# Achieving Security and Privacy in Federated Learning Systems: Survey, Research Challenges and Future Directions

*Alberto Blanco-Justicia[a], Josep Domingo-Ferrer[a], Sergio Martínez[ab], David Sánchez[a],
Adrian Flanagan[c] and Kuan Eeik Tan[c],*

[a] *Universitat Rovira i Virgili, Dept. of Computer Science and Mathematics, UNESCO Chair in Data Privacy, CYBERCAT-Center for Cybersecurity Research of Catalonia. Av. Països Catalans 26, E-43007 Tarragona, Catalonia*
{alberto.blanco,josep.domingo, sergio.martinezl, david.sanchez}@urv.cat
[b] *Universitat Oberta de Catalunya (UOC), Rbla. Poblenou 156, E-08018 Barcelona, Catalonia*
[c] *Europe Cloud Service Competence Center, Huawei Technologies Oy (Finland) Co. Ltd, Itämerenkatu 9, Helsinki, FI-00180 Finland*
{adrian.flanagan, kuan.eeik.tan}@huawei.com

## Abstract

Federated learning (FL) allows a server to learn a machine learning (ML) model across multiple decentralized clients that privately store their own training data. In contrast with centralized ML approaches, FL saves computation to the server and does not require the clients to outsource their private data to the server. However, FL is not free of issues. On the one hand, the model updates sent by the clients at each training epoch might leak information on the clients' private data. On the other hand, the model learnt by the server may be subjected to attacks by malicious clients; these security attacks might poison the model or prevent it from converging. In this paper, we first examine security and privacy attacks to FL and critically survey solutions proposed in the literature to mitigate each attack. Afterwards, we discuss the difficulty of simultaneously achieving security and privacy protection. Finally, we sketch ways to tackle this open problem and attain both security and privacy.

## 1. Introduction

Federated learning (FL) is a machine learning (ML) strategy and a related architecture by which a server learns a ML model by aggregating client-supplied local models that are trained on the clients' private data. Federated learning was presented in (McMahan et al., 2017) *as the learning task solved by a loose federation of participants devices, which are coordinated by a central server.* More in detail, in each training epoch, the server sends the current model to the clients, who return to the server a model update based on their respective private data sets; the server then aggregates the received updates and updates the central model accordingly. The process is repeated indefinitely, or while the model changes from epoch to epoch, that is, until the model converges. Therefore, FL allows building an entire ML model without sharing the clients' data (which remain in their local devices), and by leveraging the computation capabilities of the clients' devices (thereby alleviating the load at the server). A number of industries including telecommunications (Flanagan et al., 2020), internet services (McMahan et al., 2017), social networks (Li et al., 2020) or healthcare (Xu and Wang, 2019) use FL for the creation and upgrade of models in ML processes. Use cases of federated learning include text prediction for smartphone keyboards (Bonawitz et al., 2019), speech recognition in intelligent assistants (Leroy et al., 2019) or video recommendations (Ammad-Ud-Din et al., 2019).

From a privacy perspective, federated learning fulfills several design principles stated in the privacy-by-design recommendations by ENISA (Danezis et al., 2015), since (1) only meaningful data are used to update the local models; (2) the clients' raw private data never leave their devices; and (3) model updates are *aggregates* of the client's data. This makes FL a more privacy-friendly approach to data analysis than the traditional approach based on centralized data collection and centralized model training. Regarding security, secure communication channels (SSL/TLS) are assumed to be used in all client-server communications. However, FL is still vulnerable to security and privacy attacks. Regarding security, FL is vulnerable to Byzantine attacks, which aim at preventing the model from converging, and to poisoning attacks, which aim at causing convergence to a wrong model (Bhagoji et al., 2019; Kairouz et al., 2019). Regarding privacy, it has been shown (see privacy attacks in (Kairouz et al., 2019)) that client-supplied models or updates can leak certain attributes from the client's private data (attribute inference attacks). Through these attacks, a malicious server can learn whether a particular data point has been used in the training process (membership attacks) or sample data points from the distribution of the data used during the training process (reconstruction attacks). Some of these privacy attacks can even be orchestrated by other clients participating in the FL network.

In this paper, we survey the recent literature on mitigating security and privacy attacks. Our analysis includes a critical review of each solution and an empirical evaluation and comparison of the performance of security-enhancing methods. Moreover, unlike recent surveys (Kairouz et al., 2019) that deal with privacy and security issues independently, in this work we analyze the difficulties that arise when trying to satisfy both security (on the server side) and privacy (on the client side) at the same time. In this respect, we argue that the limitations of the current security- and privacy-enabling solutions make achieving both security and privacy protection an open problem. We conclude the survey by highlighting the outstanding research challenges inherent to this open problem and by sketching ways to achieve both types of protection.

The rest of paper is organized as follows. Section 2 introduces and formalizes the FL architecture. Section 3 depicts security threats to FL and surveys and empirically evaluates mitigation methods. Section 4 surveys privacy attacks to FL and reviews privacy-enhancing mechanisms. Section 5 discusses the difficulties that arise when attempting to make privacy protection for FL clients compatible with model security, and sketches solutions to simultaneously attain both types of protection. Finally, Section 6 gathers concluding remarks.

## 2. Federated learning

In a federated learning scenario (McMahan et al., 2017), a model manager initializes a learning model, such as a neural network, with parameters $\theta^0$, loss function $L$, epochs $e$, and learning rate $\rho$. Other hyper-parameters may apply, such as dropout rate, decay and momentum, but we restrict to stochastic gradient descent (SGD) (Bottou, 2010; Haykin, 2010), which is the core component on which most of today's learning algorithms rely, whether for training neural networks (Haykin, 2010), regression (Zhang, 2004), matrix factorization (Gemulla et al., 2011) or support vector machines (Zhang, 2004).

The model manager may or may not pre-train the model with available public or private data already in her possession.

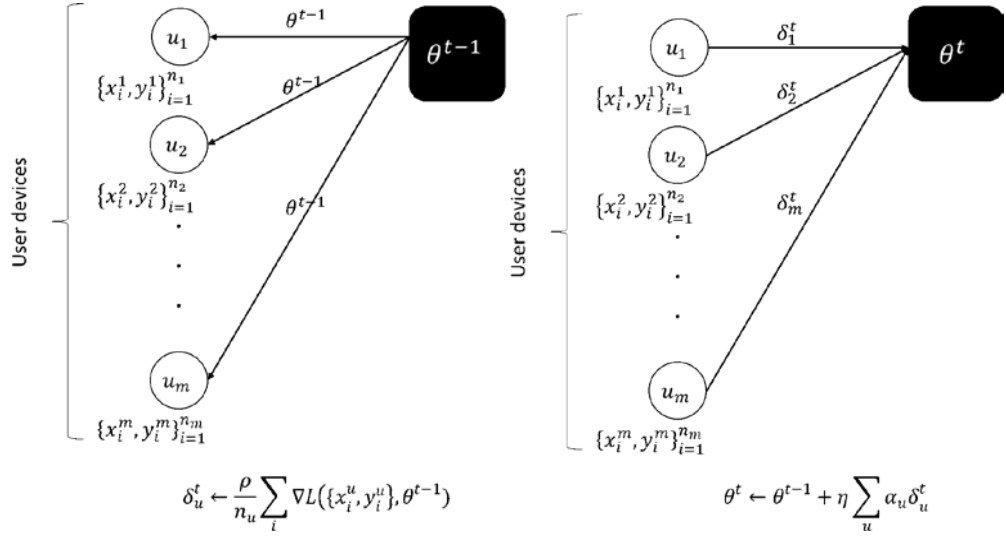

**Figure 1.** Federated learning architecture

Each of the $m$ clients, whose devices are called clients or edge devices, has access to a data set $D_u = \{x_i^u, y_i^u\}_{i=1}^{n_u}$ of size $n_u$ that she does not share with the model manager, whose device is called server. The total size of the available data is $n = \sum_{u=1}^{m} n_u$. The client devices train local models from the global model using their respective private data sets $D_u$ and send the updates $\delta_u^{t+1}$ to the model manager, who updates the global model $\theta^{t+1}$ by averaging the updates, possibly subject to a parameter $\eta$ which regulates the model substitution rate.

The typical algorithm for federated learning is formalized in Algorithm 1.

**ALGORITHM 1:** *FEDERATED LEARNING*

```
1  Initialize model parameters θ⁰ and distribute them among
   the clients
2  For every round t do
3     For every client u do
4        δᵤᵗ ← (ρ/nᵤ) Σᵢ ∇L({xᵢᵘ,yᵢᵘ},θᵗ⁻¹)
5        Collect client contributions δᵤᵗ
6     End for
```

```
7       θ^t ← θ^{t-1} + η Σ_u α_u δ_u^t
8       Distribute θ^t among clients
9   End for
```

Client selection (Line 3 of Algorithm 1) is often randomized, in such a way that only a random sample of clients is considered. Even in case no random sampling is explicitly performed, the nature of the architecture, in which edge devices are not always online, causes sampling to occur by default. The collection of local contributions (Line 5 of Algorithm 1) might be preceded by a gradient clipping procedure, which bounds the maximum and minimum possible changes to the model parameters.

Note that the server controls neither the private input of each client $\{x_i^u, y_i^u\}_{i=1}^{n_u}$ nor the updates $\delta_u^t$, which opens a window of opportunity for potentially malicious clients.

## 3. Security attacks and countermeasures

In this section, we discuss the security implications of federated learning. From Section 2, we consider a scenario where the data $D$ are split into $m$ shards $D_u$, each of them held by a client $u$. For each iteration $t$, each of the clients trains a local model $\theta_u^t$ from the global model $\theta^{t-1}$ and its local data shard. Then, they send their updates $\delta_u^t = \theta^{t-1} - \theta_u^t$ to the FL server, or model manager, who aggregates the updates to produce an updated global model $\theta^t = \theta^{t-1} + \sum_{u=1}^{m} \alpha_u \delta_u^t$, with $\alpha_u$ typically equal to $\frac{1}{m}$. Since a rational model manager has no interest in perturbing or damaging the resulting model, we assume throughout this section that the FL server is *honest* (or *honest-but-curious*), but does not have access to the client-held data shards. In contrast, clients may be *malicious* and can thus deviate from the protocol, typically by modifying either their data shard $D_u$ or their updates $\delta_u^t$. Malicious clients can act on their own or collude with other malicious clients. Any deviation from the protocol, however, is "hidden in the crowd" during the aggregation phase, as all local updates are averaged.

As a first observation, we can see that the linear aggregation function (a weighted sum) is vulnerable to manipulation by clients. If a certain malicious client had access to the other clients' updates, the malicious client would be able to replace the global model trivially. Even if assuming that a malicious client can access all other peers' updates may seem too strong, note that in the federated learning scenario (in which the global model $\theta^t$ tends to converge to optimality) the differences between client updates from time $t$ to time $t+1$ tend to 0. Therefore, a malicious client $m$ can obtain a fairly good estimate of the value of the weighted sum of the contributions by other clients by simply subtracting her own contribution from the previous global model.

The observation that a single malicious client can influence the global model opens the possibility of conducting Byzantine attacks, in which the malicious client's aim is to prevent the model from converging, or model poisoning attacks, in which the attacker's aim is to make the model misinterpret some of the inputs.

The rest of this section analyses Byzantine and poisoning attacks in detail, as well as countermeasures proposed in recent literature.

## 3.1 Byzantine attacks

In a general sense, a Byzantine fault, from the Byzantine General's Problem, refers to the problem of reaching consensus in a distributed system. Problems to reach consensus may arise because of transmission errors between clients, attackers in the communication channels who modify the transmitted information, or malicious clients (Lamport et al., 1982).

The work in (Blanchard et al., 2017) studies the resilience to Byzantine attacks of distributed implementations of stochastic gradient descent, which is the building block of FL. This includes clients that, while not being necessarily malicious, submit defective updates to the server. These defective updates may originate from software bugs or errors in communications, but they can also come from clients whose data have a legitimately different distribution from the rest of clients' data, or from malicious clients who modify their data or their updates to cause the global model not to converge.

First, the work (Blanchard et al., 2017) explores how the aggregation of updates typically takes place, namely using a weighted sum, and demonstrates that this approach is vulnerable to Byzantine attacks performed by a single (malicious) client. Any client can induce the model manager to obtain any arbitrary aggregate update, no matter whether this aggregate update is in line with updates sent by other clients or not.

**Lemma 1.** *Consider an aggregation rule $F_{lin}$ of the form $F_{lin}(\delta_1, \ldots, \delta_m) = \sum_u \alpha_u \delta_u$, where the $\alpha_u$'s are non-zero scalars, for example $\frac{1}{m}$. Let T be any vector in $\mathbb{R}^d$. A single malicious client user can cause $F_{lin}$ to always yield T.*

**Proof.** Immediate: if the Byzantine malicious client sends $\delta_m = \frac{1}{\alpha_m} T - \sum_{u=1}^{m-1} \frac{\alpha_u}{\alpha_m} \delta_u$, then $F_{lin} = T$. QED

If the malicious client has access to the updates by the rest of the clients, this attack is immediate. This is not realistic in the typical scenario, and thus the malicious client would have to estimate the values (or the sum of the values) sent by the rest of the clients. Yet, as mentioned above, if a model is close to convergence, the loss function will return small values and thus the updates of all clients will be small and similar from round $t$ to round $t + 1$. In this situation, a malicious client can estimate with high accuracy the sum of the updates of the rest of the clients from a previous round, just by subtracting its own update from the global model. This allows the malicious client to completely substitute the global model (up to some error) by creating a special local update $\delta_m^t$ that nullifies the rest of the clients' updates while boosting its own update by a factor $\frac{1}{\alpha_m}$:

$$\delta_m^t = \frac{1}{\alpha_m} T - \frac{1}{\alpha_m}(\theta^{t-1} - \delta_m^{t-1}) \xrightarrow{\alpha_u = \frac{1}{m}} \delta_m^t = m(T - \theta^{t-1} + \delta_m^{t-1})$$

A successful attack of this kind would reduce the ability of the FL server to obtain accurate outputs from the resulting model, because the global model held by the FL server does not converge. Note that the resulting model distortion survives only a few epochs after the attack is launched (possibly only one epoch). A large number of clients combined with good client selection strategies can mitigate the long-time effects of these attacks.

## 3.2 Model poisoning attacks

Model poisoning attacks intend to induce adversarial learning targets in the global model. For example, in a video recommendation scenario, a possible aim would be to include a specific video for all possible viewer histories. In the federated learning scenario, the FL server and the clients do not exchange any data save for the model updates that are obtained from training the global model on their private data. Therefore, what the adversaries alter to achieve their objectives are the model updates. Note that since the FL server does not have any information on the clients' private data, a malicious client can alter her local data shard as she sees fit.

In (Bhagoji et al., 2019) the authors explore the threat of model poisoning attacks on federated learning initiated by a single, non-colluding malicious agent whose adversarial objective is to cause the model to misclassify a single adversary-chosen input example with high confidence. The authors propose several attack strategies of increasing sophistication, from simple boosting of the malicious client's update to overcome the other clients' updates, to techniques aimed at defeating detection measures by the centralized FL server (see Section 3.3). The authors conclude that a single, highly constrained malicious client is enough to conduct poisoning attacks. The relevance of this class of attacks is limited, since misclassifying a single input example would not impact the service in any significant way. Nonetheless, we describe the attacks presented in (Bhagoji et al., 2019) because they can be generalized to other kinds of attacks which might be harmful. In the experimental Section 3.4, we use these techniques to carry out attacks that cause the global model to misclassify a whole class of examples that have some information in common.

The following assumptions are made regarding the adversary (Bhagoji et al., 2019):

- An adversary controls exactly one non-colluding, malicious client with index $m$ (the effect of malicious updates is limited).
- The data of the clients are identically and independently distributed (i.i.d.). This makes it easier to discriminate between benign and possibly malicious updates and harder to achieve attack stealth.
- The malicious client has access to a subset of the training data $D_m$ as well as to auxiliary data $D_{aux}$ drawn from the same distribution as the training and test data that are part of her adversarial objective.

These assumptions picture a very restricted adversary, which is a good starting point to evaluate how strong model poisoning attacks can become in the federated learning scenario.

A single malicious client is clearly more restricted than a group of malicious colluding clients. As described in Section 2, the global model is obtained by averaging all clients' contributions. If a substantial fraction of the clients were to collude, they could easily guide the global model to any desired objective.

Having the client-held data shards be i.i.d. limits the power of an adversary since her updates cannot deviate too much from the benign clients' updates. Thus, the way in which the malicious client poisons the model updates is constrained, which reduces the adversary's ability to affect the model.

The third assumption implies that the adversary holds some benign data, obtained in the same way as the rest of the clients, as well as auxiliary data that are altered in some way to contain some misclassifications to achieve the adversarial goal.

The goal of the adversary is to ensure that the global model learns the targeted misclassifications of the auxiliary database. The auxiliary data consist of samples $\{x_i\}_{i=1}^{n_m}$ with true labels $\{y_i\}_{i=1}^{n_m}$ that have to be misclassified as desired target classes $\{\tau_i\}_{i=1}^{n_m}$. Thus, the adversarial objective is:

$$\mathcal{A}(D_m \cup D_{aux}, \theta^t) = \max_{\theta^t} \sum_{i=1}^{n_m} \mathbf{1}[f(x_i;\theta^t) = \tau_i].$$

Note that, in contrast to Byzantine attacks, here the goal of the adversary is *not* to prevent convergence of the global model. Thus, any attack strategy used by the adversary must ensure that the global model converges to a point with good performance on the test set.

From the adversarial goal, the two challenges for the adversary are as follows:

- First, the objective represents a difficult combinatorial optimization problem. Hence, it needs to be relaxed in terms of a loss function for which automatic differentiation can be used.
- Second, the adversary does not have access to the global parameter vector $\theta^t$ for the current iteration and can only influence it through the magnitude of the update $\delta_m^t$ it provides to the server. Thus, the adversary performs the optimization over $\hat{\theta}^t$, which is an estimate of the value of $\theta^t$ based on all the information $\mathcal{I}_m^t$ available to the adversary.

The objective function for the adversary to achieve targeted model poisoning on the $t$-th iteration is

$$arg\ min_{\delta_m^t} L\big(\{x_i, \tau_i\}_{i=1}^{n_m}, \hat{\theta}^t\big), \text{s.t.}\ \hat{\theta}^t = g(\mathcal{I}_m^t),$$

where $g(\cdot)$ is an estimator. The estimate $\hat{\theta}^t = \theta^{t-1} + \alpha_m \delta_m^t$ implies that the malicious client ignores the updates from the other agents but accounts for scaling at aggregation. This assumption is enough to ensure the attack works in practice.

*Explicit boosting attack*. In (Bhagoji et al., 2019), the authors describe first a trivial attack based on explicit boosting of the local malicious update, which has been trained according to the target selected by the adversary. The adversary could directly optimize the adversarial objective $L\big(\{x_i, \tau_i\}_{i=1}^{n_m}, \hat{\theta}^t\big)$, with $\hat{\theta}^t = \theta^{t-1} + \alpha_m \delta_m^t$, where $\tau_i$ are the modified class labels according to the target defined by the adversary. However, this setup implies that the optimizer has to account for the scaling factor $\alpha_m$ implicitly.

A trivial way to overcome the scaling factor is to explicitly boost the local malicious update, that is, to scale it by some factor. By mimicking a benign agent, the malicious agent can run $E_m$ steps of a gradient-based optimizer starting from $\theta^{t-1}$ to obtain $\tilde{\theta}^t$, which minimizes the loss over $\{x_i, \tau_i\}_{i=1}^{n_m}$. The malicious agent then obtains an initial update $\tilde{\delta}_m^t = \tilde{\theta}^t - \theta^{t-1}$. However, since the malicious agent's update tries to ensure that the model learns different labels from the true labels for the data of its choice ($D_{aux}$), it has to overcome the effect of scaling, which would otherwise mostly nullify the desired classification outcomes. This happens because the learning objective for all the other agents is very different from that of the malicious agent, especially in the i.i.d. case. The final update sent back by the malicious agent is then $\delta_m^t = \lambda \tilde{\delta}_m^t$, where $\lambda$ is the factor by which the malicious agent boosts the initial update. Note that with $\hat{\theta}^t = \theta^{t-1} + \alpha_m \delta_m^t$ and $\lambda = 1/\alpha_m$ then $\hat{\theta}^t = \tilde{\theta}^t$, implying that if the estimation was exact, the global update vector should now satisfy the malicious agent's objective. This follows the same rationale as the Byzantine attack.

Note that whereas this attack might be effective at introducing the malicious target into the global model, it does not take into account any detection strategies put in place by the FL server. Since the adversary target greatly differs from the benign clients, an accuracy test as we describe in Section 3.3 could be enough to flag the malicious updates as suspicious. Statistics on the update magnitudes might also indicate notable differences between the malicious and the benign updates.

*Stealthy boosting attack*. In order to bypass the potential detection mechanisms put in place by the FL server, the authors of (Bhagoji et al., 2019) propose a refined variant of the explicit boosting attack. In addition to boosting the malicious updates by some scalar $\lambda$, the malicious client can add more loss terms to the learning target to maintain acceptable levels of accuracy, thereby bypassing accuracy and validation loss checking controls, and at the same time obtaining statistics of the update magnitudes similar to those of the rest of the clients.

First, in order to improve the accuracy on validation data, the adversary adds the training loss over the malicious agent's local benign data shard $D_m(L(D_m, \theta^t))$ to the objective. Since the training data should be i.i.d. to the validation data, this will ensure that the malicious agent's update is similar to that of a benign agent in terms of validation loss and accuracy and will make it challenging for the server to flag the malicious update as anomalous.

Second, the adversary needs to ensure that her update is as close as possible to the benign agents' updates in the appropriate distance metric. As we describe in Section 3.3, the model manager can use distance-based metrics (e.g. $\ell_2$) to detect potentially malicious updates. However, the adversary does not have access to the updates for the current time step $t$ that are generated by the other clients, but can obtain them from the previous global model by subtracting her own scaled contribution to obtain the average benign update $\tilde{\delta}_{ben}^{t-1} = \sum_{i \in [1, m-1]} \alpha_i \delta_i^{t-1}$. The malicious client then constrains $\delta_m^t$ to the average benign update, by adding a loss term to its objective in the form of $\phi \|\delta_m^t - \tilde{\delta}_{ben}^{t-1}\|_2$, where $\phi$ is an adversary-defined parameter. The addition of such training loss term could still not be sufficient to ensure that the malicious weight update is close to that of the benign agents since there could be multiple local minima with similar loss values.

Overall, the adversarial objective then becomes:
$$argmin_{\delta_m^t} \lambda L(\{x_i, \tau_i\}_{i=1}^{n_m}, \hat{\theta}^t) + L(D_m, \theta^t) + \phi \|\delta_m^t - \tilde{\delta}_{ben}^{t-1}\|_2$$

Note that for the training loss, the optimization is just performed with respect to $\theta_m^t = \theta^{t-1} + \delta_m^t$, as a benign agent would do. By using explicit boosting, $\hat{\theta}^t$ is replaced by $\theta_m^t$ as well, so that only the portion of the loss corresponding to the malicious objective is boosted by a factor $\lambda$.

## 3.3 Countermeasures to security attacks

The FL server is not completely oblivious to the potential misbehaviors of clients and, in the general case, it has access to the individual updates by clients. When this is the case, as anticipated in the previous section, the FL server can analyze both the updates and the impact of individual updates on the global model to identify misbehaving clients. The following types of approaches to detect attacks exist:

- *Detection of malicious clients via model metrics* (Chen et al., 2017). The FL server can reconstruct the individual updated models for every client $u$ as $\theta_u^t = \theta^{t-1} + \frac{\eta}{m}\delta_u^t$ and compare the model performance metrics, such as accuracy or loss, against a validation data set with respect to the model obtained by aggregating all updates except that of client $u$. The FL server can mark as anomalous and possibly discard any client updates that degrade the model performance according to some rule or threshold. Note that the FL server requires access to a validation data set, and this is not always feasible in the federated learning scenario.

- *Detection of malicious clients via update statistics.* An alternative (or complementary) approach for the FL server is to observe the statistics of the magnitudes of the updates (Yin et al., 2018). As a model converges, i.e. its loss function is minimized, the update magnitudes tend to 0. A trivial approach to check whether clients are well behaved is to compute the distance (e.g. $\ell_2$) of their updates to the $\vec{0}$ vector and mark as anomalous and/or discard those updates whose distance is out of a given range. A possible candidate for this range would be the interquartile range of all update distances times a fixed parameter. However, since the condition that updates are close to 0 only holds for models that are close to convergence, this approach is not suitable for the whole lifetime of the model. An alternative approach is to compute the centroid $\hat{\delta}^t$ of the updates submitted by the different clients, i.e., their coordinate-wise average, and compute the distances of all individual updates to $\hat{\delta}^t$. Again, updates at distances out of a certain range would be considered anomalous updates. Similarly, the FL server could compute how much do the distributions of distances in successive FL iterations change, for example using the Kullback-Leibler divergence metric. In a scenario with colluding malicious clients, these might have enough influence on the computed centroid $\hat{\delta}^t$ to render the previous countermeasures ineffective. In such cases, the FL server might compute the centroid as a median rather than as an average. The median is more robust in front of outlying updates submitted by malicious clients. Note that the previous approaches add a significant computational cost to the aggregation function.

- *Krum aggregation method*. The authors of (Blanchard et al., 2017) propose an aggregation function that is resilient against $f$ malicious peers. To that end, they combine the intuitions of the majority-based and $\ell_2$-based methods and choose the local update $\delta_u^t$ that is closest to its $m - f - 2$ neighbors. That is, the one that minimizes the sum of squared distances to its $m - f - 2$ closest vectors. They call their aggregation function Krum.

    **Krum aggregation rule** $Kr(\delta_1, \dots, \delta_m)$: For any $i \neq j$, denoted by $i \to j$ the fact that $\delta_j$ belongs to the $m - f - 2$ closest vectors to $\delta_i$. Then, for each peer user $i$, the score $s(i) = \sum_{i \to j} \|\delta_i - \delta_j\|_2^2$ is computed, where the sum runs over the $m - f - 2$ closest vectors to $\delta_i$. Then, $Kr(\delta_1, \dots, \delta_m) = \delta_k \ s.t. \forall i \ s(k) \leq s(i)$.

    Assuming $2f + 2 < m$, they show that Krum is resilient against $f$ malicious peers and that the corresponding machine learning scheme converges. An important advantage of Krum is its (local) time complexity $O(m^2 \cdot d)$, which is linear in the dimension $d$ of the updates.

- *Coordinate-wise median*. In (Yin et al., 2018), an aggregation function is proposed that selects the coordinate-wise median instead of the coordinate-wise average. Thus, $\theta^t = \theta^{t-1} + cmed(\{\delta_u^t\}_{u=1}^m)$, where $cmed(\cdot)$ returns the median for every coordinate of the local updates. Since the median is a more robust statistic than the mean (i.e. it is less influenced by outliers), the obtained global model is less influenced by potential malicious peers.

- *Coordinate-wise trimmed mean*. Also in (Yin et al., 2018), a second aggregation function is proposed, called coordinate-wise trimmed mean. In this case, the contributions $\delta_u$ are first ordered, then a fraction $\beta$ of the smallest and a fraction $\beta$ of the largest contributions are removed. The global model is computed as $\theta^t = \theta^{t-1} + \frac{1}{(1-2\beta)m}\sum_{r \in R} \delta_r^t$ where the set $R$ contains the remaining contributions.

Table 1 summarizes, characterizes and compares the security attacks and countermeasures discussed so far.

Table 1. Security attack methods on FL along with their characteristics and possible countermeasures

| Security attack | Impact | Requirements | Countermeasures | Result of countermeasures | Cost of countermeasures |
|---|---|---|---|---|---|
| Byzantine attacks (Blanchard et al., 2017) | The model does not converge | Estimate the values (or the sum of the values) sent by the rest of the clients | • Large number of epochs<br>• Large number of clients<br>• Good client selection<br>• Detection of malicious clients via model metrics; it requires a test data set (Chen et al., 2017)<br>• Detection of malicious clients via update statistics (Yin et al., 2018)<br>• Krum aggregation method (Blanchard et al., 2017)<br>• Coordinate-wise median (Yin et al., 2018)<br>• Coordinate-wise trimmed mean (Yin et al., 2018) | -<br><br>-<br><br>-<br>List of possible attackers<br><br><br><br>List of possible attackers<br><br><br>Robust aggregate<br><br>Robust aggregate<br><br>Robust aggregate | Low<br><br>Low<br><br>Low<br>High<br><br><br><br>Medium<br><br><br>Low<br><br>Medium<br><br>Medium |
| Model poisoning attack through explicit boosting (Bhagoji et al., 2019) | Induce adversarial learning targets in the global model | Access to a subset of the training data and auxiliary data with the same distribution | | | |
| Model poisoning attack through stealthy boosting (Bhagoji et al., 2019) | Induce adversarial learning targets in the global model maintaining acceptable levels of accuracy | Access to a subset of the training data and auxiliary data with the same distribution | | | |

## 3.4 Empirical analysis

For the attacks described in Sections 3.1 and 3.2 to succeed, the attacker needs to overwhelm the influence of the rest of the clients in the aggregation function. The attacker can achieve this in three ways: the first is for the attacker to boost her own updates by a scaling factor $\beta$; the second is to collude with other malicious clients, either genuine different clients or several fake accounts created by the attacker; the third is to combine both approaches.

We report below experimental results on these attack strategies, and we test the protection offered by detection mechanisms described in Section 3.3.

Our experiments use the MNIST handwritten digits data set. The base global model is a neural network where all convolutional layers and the dense first layer use the ReLU activation function. The second dense layer, which is the output layer, uses softmax. The loss function is categorical *crossentropy*. The model achieves 98.89% validation accuracy after 10 epochs and a batch size of 32.

The federated version of the base scenario uses the same architecture and parameters. The data set is evenly distributed among 10 users, following the example of (Bhagoji et al., 2019). (Since 10 is a very small number, in the next experiment we will increase it for the sake of realism). The model substitution rate $\eta$ is set to 0.25, meaning that in each epoch the model keeps information about past training epochs, which makes the training process more stable because there are no abrupt changes in the model performance from epoch to epoch. To compensate for the low substitution rate, the training process needs to be longer. The training process is carried out for 20 global epochs, in which all clients participate. Local updates are computed after 2 epochs of local training, with a batch size 32. The resulting validation accuracy is 98.54%.

Next, we introduce a malicious client, whose objective is to introduce a backdoor in the model so that it classifies the digit '4' as class '7'. We test how effective the attack is for boosting values $\beta$ from 2 to 12. Results for the extreme values of $\beta$ ,2 and 12 are given in the graphs below, whose lines must be read as follows:

- In blue, the global accuracy of the model.
- In orange, the accuracy of the model without considering the class '4', which is the one subjected to the attack.

- In green, the proportion of '4' digits that are misclassified as '7'. This quantifies the success of the attack.

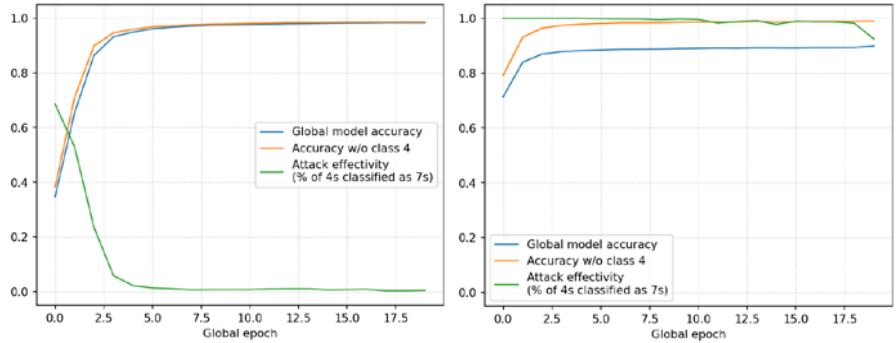

**Figure 2.** Model poisoning attack with explicit boosting. Network with 10 users, among whom one attacker. Results for boosting values 2 (left) and 12 (right).

While scaling factors smaller than the number of participants (lower than 10) are capable of introducing the backdoor, only boosting factors equal to the number of participants or greater achieve attack effectiveness (the green line) significant with respect to the global model accuracy (the blue line).

The above experimental scenario is a worst-case setting, inspired by the work of (Bhagoji et al., 2019). However, this scenario is not realistic in production systems, since the attacker is allowed to participate in all training epochs. A more realistic scenario is one in which only a small fraction of clients participates in each global training epoch.

To simulate this, in the following experiment, we increase the total number of peers to 1,000, and select a random sample of clients to participate in each of the training epochs. We repeat this experiment for sample sizes 1% (realistic scenario) and 10% (far beyond the realistic scenario) of the client pool. The number of global training epochs is increased to 100 to allow for a diverse selection of clients, especially in the 1% scenario. The rest of parameters stay unchanged, including the single malicious client's objectives, who will boost her own updates by a factor equal to the number of clients participating in each training epoch (10 and 100, respectively).

The figures below show the results of this configuration, where the meaning of the lines is the same as in the above graphs.

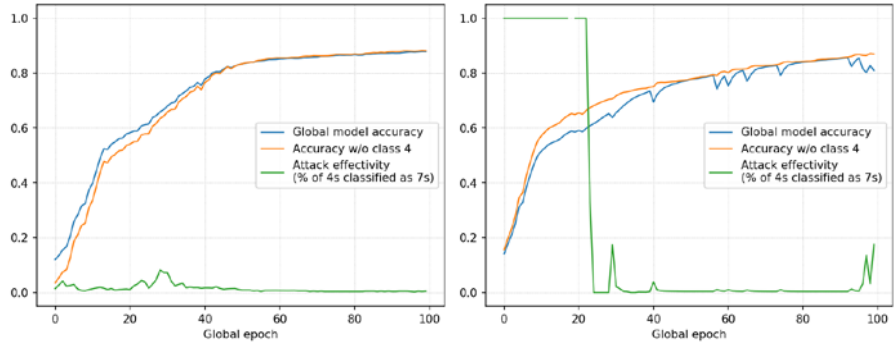

**Figure 3.** Model poisoning attack with explicit boosting. Network with 1,000 users among whom one attacker. Left, boosting value 10. Right, boosting value 100.

We can see that:

- While the malicious client is able to introduce the backdoor, it performs the attack with very low effectiveness and only for a few epochs after having been selected.
- As previously said, there are statistical mechanisms to detect malicious updates, which we evaluate below.

From now on, all experiments will be conducted in this scenario, with 1,000 client peers where 1% of the client base participate in each global training epoch.

Next, we implement several protection mechanisms at the aggregation phase among those presented in Section 3.3, including outlier detection by either update statistics or from the global model accuracy, median aggregation, and the Krum aggregation rule.

The initial scenario to test the effectiveness of the countermeasures follows the configuration described above, that is, a network with 1,000 clients of which 1% participate at each global training epoch. A single malicious client attempts to misclassify class '4' as '7'. As shown in Figure 4 to Figure 7, all considered countermeasures reduce the already low success chance of the attacks. Note that the first 20 global training epochs for all experiments in this section and the next one might give the impression of a higher success rate than expected. However, this effect can be explained by the low global accuracy at the initial stages of training. Further into the training process, the effectiveness of the attack falls close to 0 for all tested countermeasures.

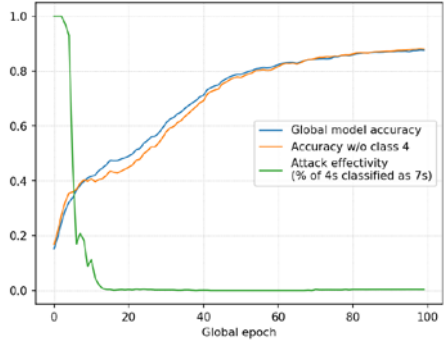
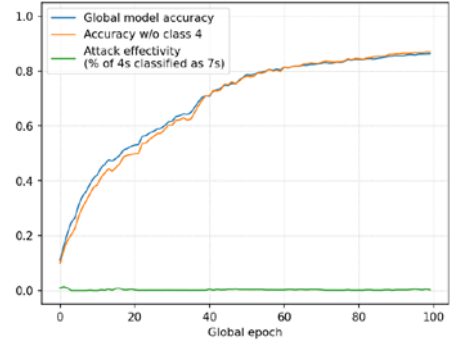

**Figure 4.** Distance detection          **Figure 5.** Model accuracy detection

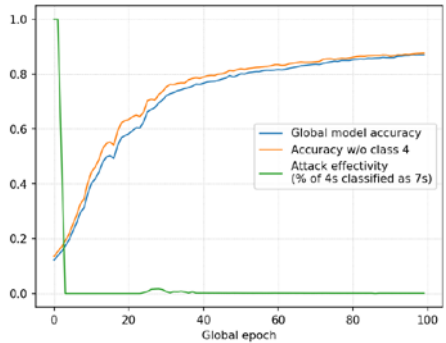
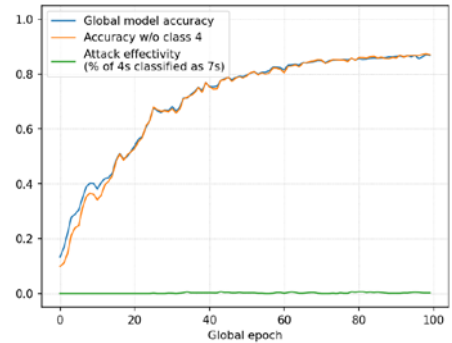

**Figure 6.** Median aggregation          **Figure 7.** Krum aggregation

Next, we test the performance of federated learning when subjected to poisoning attacks by coalitions of malicious clients, all with the same objective. The coalitions grow from 10% to 40% of the total number of clients. As in the previous setting, only 1% of the clients participate in each training epoch. Figures 8 to 11 show the effectiveness of the poisoning attacks when no countermeasures are applied. The boosting factors β of the malicious clients have been set to 2, from at least 10 in the previous experiments, and so each of the individual attackers is harder to be identified. From the experiments, in a scenario with 10% or more malicious colluding clients, the model will be successfully attacked even when only 1% of the clients participate in each global training epoch.

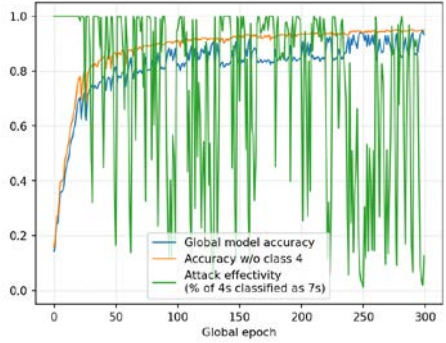

**Figure 8.** 10% malicious peers

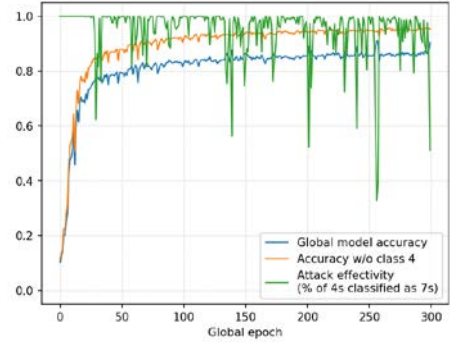

**Figure 9.** 20% malicious peers

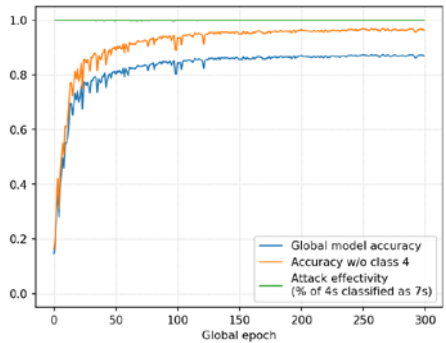

**Figure 10.** 30% malicious peers

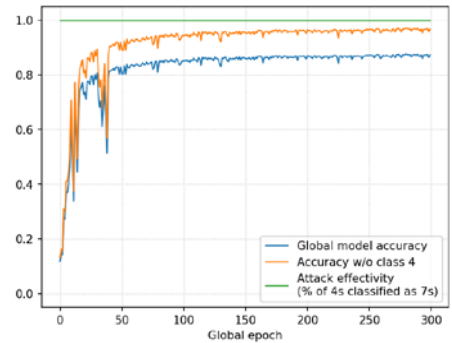

**Figure 11.** 40% malicious peers

Now, we test the same protection mechanisms of Section 3.3 against colluding attackers. The countermeasures are tested against Byzantine and poisoning attacks executed on an average case of a coalition of 20% (Figure 12 to Figure 15) and on the extreme case of a coalition of 40% (Figure 16 to Figure 19) of the client base. The rest of the testing parameters remains unchanged.

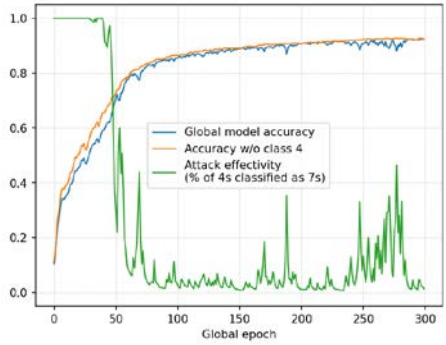

**Figure 12.** 20% coalition. Distance detection

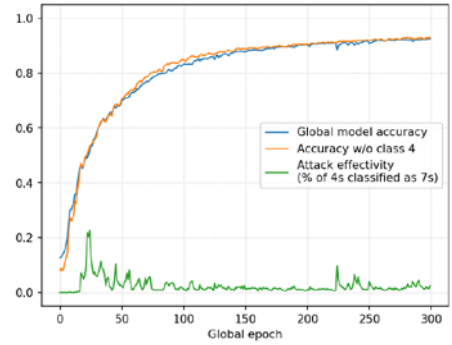

**Figure 13.** 20% coalition. Model accuracy detection

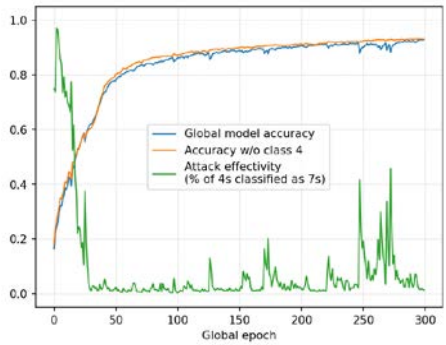

**Figure 14.** 20% coalition. Median aggregation

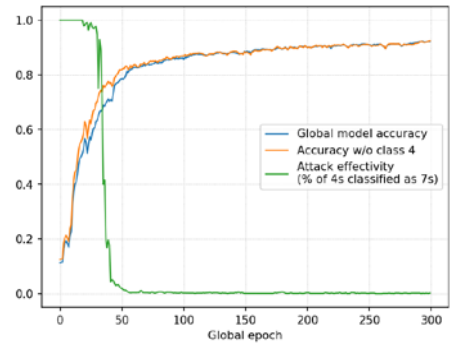

**Figure 15.** 20% coalition. Krum aggregation

According to our experiments, for coalitions up to 20% of client base, the detection and discarding of outliers based on the distance to the average update offers the worst results (see Figure 12). The other countermeasures offer good results. In particular, the median aggregation rule (Figure 13) is low in complexity and offers a performance similar to model accuracy detection (Figure 14) and Krum aggregation (Figure 15) countermeasures.

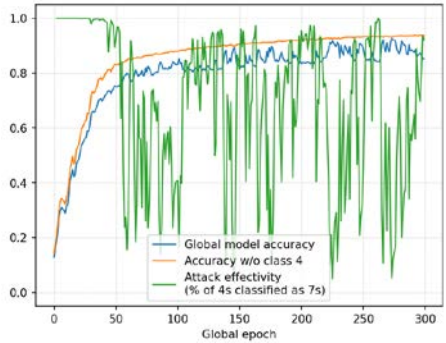

**Figure 16.** 40% coalition. Distance detection

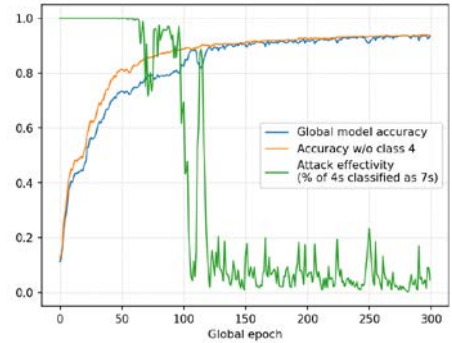

**Figure 17.** 40% coalition. Model accuracy detection

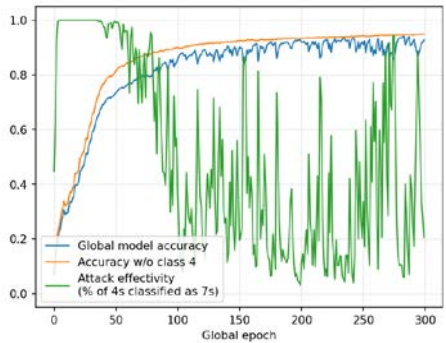

**Figure 18.** 40% coalition. Median aggregation

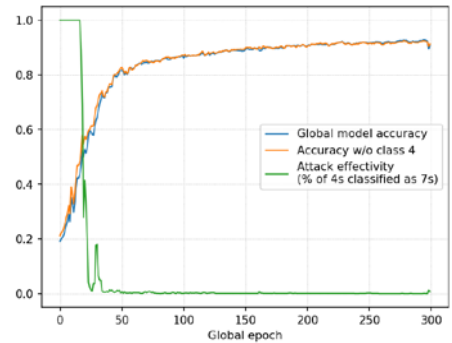

**Figure 19.** 40% coalition. Krum aggregation

In the extreme case of coalitions up to 40% of the client base, the detection and discarding of outliers based on the distance to the average update and the median aggregation rule are low in complexity but offer bad results (Figure 16 and Figure 18). Outlier detection based on the model accuracy seems to offer the highest protection against Byzantine and poisoning attacks (see Figure 17). However, the FL server needs to have a test data set, which may not always be at hand. Moreover, time and space complexity for this countermeasure are also much higher than for the rest of methods, since the model must be evaluated for each individual client update. Finally, as shown in Figure 19, the Krum aggregation countermeasure offers the best results, which are very similar to those obtained by the model accuracy aggregation rule but at a much lower computational cost and without requiring a test data set.

*3.5 Conclusions on security attacks*

Our experiments show that on very large communities of clients where only a small fraction of clients is randomly selected to contribute at each global training epoch, an attack by a single malicious client has little to no effect on the global model metrics. This result may not necessarily apply to all classification or regression tasks (for example to tasks with many more labels than the 10 labels of the MNIST data set), so that the use of countermeasures such as the ones introduced in Section 3.3 and tested in Section 3.4 should be considered.

Among the discussed and tested countermeasures, the mechanism based on computing the global model accuracy and the Krum aggregation rule offered the best results under attacks by large coalitions. The drawbacks of the global model accuracy detection mechanism are that it requires a test data set and that it is computationally expensive. The Krum aggregation rule seems to be of particular interest, given its effectiveness at discarding potentially malicious updates, its low complexity and its easy implementation and application.

Note that all these countermeasures require the FL server to access individual updates from the clients. This fact restricts the possible measures to protect clients involved in FL systems from privacy attacks described in the next Section 4; in particular, the security countermeasures of Section 3.3 are incompatible with privacy protection based on update aggregation.

## 4. Privacy attacks and defenses

Federated learning intrinsically protects the data stored on each device by sharing model updates, e.g., gradient information, instead of the original data. However, model updates, which are based on original data, can reveal sensitive information. In the sequel, we characterize the adversaries that can conduct privacy attacks against FL.

In an FL algorithm, at each iteration $t$ of the training process the adversary downloads the current global model, calculates gradient updates, and sends her own updates to the FL server. The adversary saves the current global model parameters $W_t$. The difference between the consecutive global models, $\Delta W_t = W_t - W_{t-1} = \sum_k \Delta W_t^k$, is equal to the aggregated updates from all participants; hence, $\Delta W_t - \Delta W_t^{adv}$ are the aggregated updates from all participants other than the adversary.

With respect to their role, adversaries can be classified into four types:

- *Honest-but-curious FL server*. A curious FL server receives updates $W_i^t$ from each participant over time and uses $W_i^t$ to infer information about the private data set of individual clients.
- *Malicious FL server*. Such a server can perform powerful attacks because it can also control the view of each client on the global model. In this way, a malicious FL server can extract additional information about the private data set of a client.
- *Honest-but-curious client*. An adversarial honest-but-curious client *i* can only observe the global parameters over time, $W^t$, and she can use the successive parameters of the model to infer information about the private data of other clients.
- *Malicious client*. An adversarial malicious client *i* can obtain the aggregated updates from all other clients and can craft her own adversarial parameter updates $W_i^t$ in order to get as much information as possible about the private data of other clients.

Privacy attacks to FL can be classified into two fundamental and related categories (Melis et al., 2019):

- *Membership inference attacks*. They consist in determining whether an individual data record was in the training data set. The ability of an adversary to infer the presence of a specific record in the input data training constitutes an immediate privacy threat if the training data are private or sensitive.
- *Properties of training data inference attacks*. In FL, the distribution of individuals belonging to different classes may differ among the various private data sets. This attack aims at inferring properties of a class: for example, for facial recognition models, if a class corresponds to a certain individual, the attacker could infer that the individual wears glasses (Fredrikson et al., 2015).

In the following section, we survey inference attacks targeted at disclosing private clients' data. Afterwards, we discuss privacy-enhancing measures that can be employed by the clients to protect their own data.

### 4.1 Attacks from the embedding layer

When dealing with nominal data with a discrete and sparse input space (e.g., natural-language text), deep learning architectures usually employ an embedding as input layer in order to manage a vector representation with a lower dimension. Let vocabulary V be the set of all words. The embedding process maps each word in the training data to a word-embedding vector via an embedding matrix $W_{emb} \in \mathbb{R}^{|V| \times d}$, being $|V|$ the size of the vocabulary and $d$ the dimensionality of the vector.

During the training process, the embedding matrix is treated as a set of parameters of the model and it is collaboratively optimized. The gradient of the embedding layer is sparse with respect to the input words: given a batch of text, the embedding is updated only with the words that appear in the batch. The gradients of the other words are zeros. Therefore, the words that belong to the training inputs offered by the honest participants in the collaborative learning might be revealed by the non-zero gradients.

### 4.2 Attacks from the gradients

In deep learning models, gradients are computed by back-propagating the loss through the entire network from the last to the first layer. The features of the layer and the error from the previous layer are used to compute the gradients of a layer. In the case of fully connected sequential layers $h_l, h_{l+1}$ ($h_{l+1} = W_l \cdot h_l$, where $W_l$ is the weight matrix), the gradient of error $E$ with respect to $W_l$ is computed as $\frac{\partial E}{\partial W_l} = \frac{\partial E}{\partial h_{l+1}} \cdot h_l$. The gradients of $W_l$ are the inner products of the error from the previous layer and the features $h_l$. Feature values might be inferred through observations of the gradient updates. These features are in turn based on the clients' private input data in the training process.

### *4.3 Passive and active property inference*

A passive attack occurs when the adversary infers properties from the learning process without modifying the model. For example, in the passive attack proposed in (Melis et al., 2019), the adversary uses auxiliary data consisting of points that have the property of interest and points that do not have the property. These data points need to be sampled from the same class as the target participant's data, but otherwise can be unrelated. In this attack, the adversary could leverage the global model of each iteration to create aggregated updates based on the data with the property, and updates based on the data without the property. This produces labeled samples, which enable the adversary to train a binary batch classifier that determines if the observed updates are based on the data with or without the property.

An active adversary can perform a more powerful attack by altering the learning procedure. An active adversary participates in the training process by influencing the target model in order to obtain more information about the training set. In this setting, the FL server or a curious client can construct adversarial parameter updates for an active inference attack.

Specifically, the adversary can extend his local copy of the collaboratively trained model with an augmented property classifier connected to the last layer. He trains this model in order to recognize batch properties while simultaneously performing well on the main task. The adversary uploads the updates based on her joint loss during the collaborative training process, thereby causing the global model to learn separable representations for the data with and without the property, with the gradients being separable too and enabling the adversary to infer if the training data has the property. The only difference with the passive attack is that this adversary performs additional local computations and uploads the altered values into the collaborative learning procedure.

## 4.4 Generative adversarial networks (GAN)

More powerful attacks than those seen so far can be performed through Generative Adversarial Networks (GAN). The Deep Learning community has recently proposed GANs (Goodfellow et al., 2014; Radford et al., 2016; Salimans et al., 2016), which are still being intensively developed (Hitja et al., 2017). The objective of GANs is to generate similar samples with the same distribution to those in the training data. It must be highlighted that GANs generate these samples without having access to the original samples. The GAN learn the distribution of the data through the deep neural network.

The GAN procedure pits a discriminative deep learning network against a generative deep learning network in a game (in the sense of game theory). In the original paper (Goodfellow et al., 2014), the data used in the examples consist of data sets that contain images. In this case, the network is trained to discriminate between images from the original data set and the images generated by the GAN. However, GAN attacks can be extended to other types of data, such as demographic records.

In a first step, the generative network is initialized with random values. Then, at each iteration the GAN is trained to imitate the images included in the training data of the discriminative network. The GAN procedure solves the following optimization problem:

$$\min_{W_G} \max_{W_D} \sum_{i=1}^{n_+} \log f(x_i; W_D) + \sum_{j=1}^{n_-} \log(1 - f(g(z_j; W_G); W_D)),$$

where $\{x_i\}_{i=1}^{n_+}$ is a data set of real images of size $n_+$, $f(x_i; W_D)$ is a discriminative deep neural network that classifies input images as real or fake, and $W_D$ denotes its parameters. $\{z_j\}_{j=1}^{n_-}$ is a set containing randomly generated values of size $n_-$, $g(z_j; W_G)$ is a generative deep neural network with parameters $W_G$, which from a random input generates an image.

The training procedure works as follows. First the gradient on $W_D$ is computed to maximize the performance of the discriminative deep neural network. Hence $f(x_i; W_D)$ is able to distinguish between samples from the original data, i.e., $x_i$, and samples generated from the generative structure, i.e., $x_j^{fake} = g(z_j; W_G)$. Then, the gradients on $W_G$ are computed. In this way, the samples generated from $x_j^{fake} = g(z_j; W_G)$ imitate the original data (without being copies of them). At the iteration where the discriminative network cannot differentiate between samples from the original data set and the samples generated by the generative network, the procedure ends.

*Client-side GAN-based attacks*. This kind of attacks exploits the real-time nature of the training process: a possible adversary can train a GAN to generate prototypical samples of the targeted training set and, thus, compromise the privacy of the training set owner (Hitja et al., 2017). The samples generated as output of the GAN seek to have the same distribution as the input data used during the training process. In addition and interestingly, the authors in (Hitja et al., 2017) demonstrate that protecting the shared parameters that constitute the model by applying record-level differential privacy noise, as suggested in previous work (Shokri and Shmatikov, 2015), is ineffective (i.e., record-level DP does not protect against GAN-based attacks). This issue is discussed in Section 4.5.

The attack described in (Hitja et al., 2017) relies on an active insider. It works in a white-box access model where the attacker can see and use internal parameters of the model. The adversary participates as an honest client in the federated deep learning protocol, but he tries to extract information about a class of data he does not own (the target data is owned by the victim client). In the active attack, the adversary will also influence the learning process to force a victim into releasing further details about the targeted class.

Client-side GAN-based attacks have, however, three main limitations: i) they require changing the architecture of the distributed model to introduce adversarial influence in the learning process; ii) the adversarial influence introduced by the malicious client can become insignificant after many iterations of the process; iii) the attack can only imitate the input data for training rather than the exact samples from the victim (Melis et al., 2019).

*FL server-side GAN-based attacks*. In order to address the drawbacks of client-side GAN-based attacks, the authors of (Wang et al., 2019) propose a multi-task GAN for

Auxiliary Identification (named mGAN-AI) that operates on the FL server side and does not affect the learning process because it is invisible for the FL procedure. The attack improvements are based on performing additional tasks during the training process of the GAN. The improvements enhance the quality of the generated samples without compromising the collaborative learning process and without modifying the shared model, thus achieving an invisible attack. In addition, the fact that the FL server is able to discriminate the identity of the clients enables client-level retrieval of private data.

The server-side mGAN-AI attack works as follows. Assume $N$ clients where the victim $v$ is the client whose data would be reconstructed by the malicious FL server. Following the federated learning procedure, at the $t$-th iteration, the malicious FL server sends the current shared model $M_t$ to each of the $N$ clients and then receives the updates $u_t^1, u_t^2, ..., u_t^N$, which have been trained by each client using as input their respective private data. Specifically, $u_t^v$ denotes the update $u$ from the victim $v$ at time $t$. To reconstruct the victim's private data, the attack uses a variant of GAN with a multi-task discriminator, which simultaneously discriminates category, reality, and the client identity of input samples. Note that the architecture of the discriminator is the same as the shared model regardless of the output layer. Then, the FL server can perform a passive attack by aggregating the update $u_t^v$ of the victim $v$ to the shared model in this iteration and updating the discriminator, or it can perform an active attack by aggregating the update $u_t^v$ to the discriminator directly. Moreover, the FL server calculates representatives of each client to control their identities. Then, the FL server trains the updated discriminator using the samples produced by the generator as inputs. Correspondingly, the generator is updated to approximate the victim's data.

### 4.5 Privacy defenses

Privacy attacks can break the privacy of a client versus other (malicious) clients or the FL server. The attacks performed by the FL server are especially effective, since the server centralizes client updates and has more control on the FL process (particularly, on the configuration of the global model). Also, it is not unrealistic that the FL server may know the identity of a client (e.g., if the server is the service provider of a telecommunication company); in this way, a record with private data inferred on a specific client could be associated to a real identity.

Clients can, however, proactively implement privacy defenses to thwart privacy attacks (especially those conducted by the FL server). In the literature, we can find solutions that use secure multiparty computation (SMC) to securely aggregate updates so that the server can only obtain the clear aggregated updates, but not individual ones. On the other hand, some authors employ differential privacy (DP) to distort client updates locally, so that attack inferences are no longer unequivocal. In the following we review these two approaches and discuss their pros and cons.

*Secure multiparty computation*. Cryptographic techniques have been used to prevent privacy disclosure of client data in federated learning. The best-known approach is that of Google, which proposes to use a secure aggregation protocol (Bonawitz et al., 2017) where the updates from individual clients' devices are securely aggregated by leveraging SMC to compute weighted averages of model parameters. The FL server can decrypt the average update only if a sufficient number of clients (for example 100 or 1000) have participated in the secure aggregation. This is feasible thanks to the client updates being sent via additive secret sharing. No individual update is disclosed before averaging, but this is not a problem because the FL server only needs the average update to update the global model. As a result, the private data of clients participating in the FL protocol are protected.

The main advantage of this approach is that SMC is a lossless method; it therefore retains the accuracy of the learned model while keeping the data private. On the negative side, SMC protocols incur significant extra communication cost among clients. Such a cost (which ought to be added to the cost of the local training) may be unaffordable for some client devices and networks. Also, the secure aggregation of updates requires the clients to communicate with each other; this may be difficult in highly dynamic networks (such as mobile networks) in which client devices are not always on-line or reachable or in which the physical communication channels may not be reliable. Finally, as it will be discussed in Section 5, SMC protection has a very negative side effect for security: due to the secure aggregation of updates, it disables any countermeasures that the FL server may implement to mitigate Byzantine and poisoning attacks against the model. Since SMC-based protection hides the individual updates provided by the clients, countermeasures to security attacks are ineffective and attacks become undetectable to the server.

*Differential privacy*. Locally distorting client updates by adding noise with a distribution that offers ε-differential privacy is a very common approach to enhance privacy in FL (Bhowmick et al., 2019; Differential Privacy Team, 2017; McMahan and Andrew, 2018; Triastcyn and Faltings, 2019; Wei et al., 2020).

Differential privacy (DP) was originally proposed for interactive statistical queries to a database (Dwork, 2011). Formally, a randomized algorithm $M$ with domain $\Lambda^{|\mathcal{X}|}$ is $\epsilon$-differentially private if for all $S \subseteq Range(M)$ and for all datasets $x, y \in \Lambda^{|\mathcal{X}|}$ that differ in a single record, it holds that
$$Pr[M(x) \in S] \leq e^\epsilon Pr[M(y) \in S].$$
This implies that the result of algorithm $M$ is not affected (up to some probability) by the inclusion, removal or modification of a single record. If each record stores the data of a single individual, differential privacy offers a very robust privacy guarantee: the outputs of differentially private algorithms are not (very) affected by the presence or absence of any individual's data in the input data set. Differential privacy is often achieved by adding controlled noise to the input data or to the output of algorithm $M$. Specifically, the noise distribution depends on the $\varepsilon$ parameter (which should be below 1 to offer robust protection (Dwork, 2011)) and the sensitivity of $M$ to the presence or absence of any single record.

When applied to the FL setting, DP distorts client updates so that the presence or absence of any particular record in a client's private data does not significantly influence the update sent by the client. Therefore, no unequivocal inferences can be made by the FL server on the clients' data based on individual updates.

Compared to SMC, DP has the advantage that the required noise can be efficiently added at the client side. Also, neither changes in the protocol nor extra communication are needed. Finally, DP, when applied properly, offers a strong privacy guarantee.

The main drawback of DP is that the distortion introduced by the noise added to the client updates significantly affects the accuracy of the learned model (Domingo-Ferrer et al., to appear). For instance, in (Wei et al., 2020), the authors show that using DP in FL significantly decreases the accuracy of the model, even for very large values of $\varepsilon$ (between 30-100). Notice that the privacy guarantees of DP for such large values of $\varepsilon$ are so weak that they would not prevent privacy attacks. Moreover, since model updates are protected in each epoch, and in successive epochs they are computed on the same (or, at least, not completely disjoint) client data, sequential composition applies. The sequential composition property of DP states that if a data set collected at time $t_1$ is DP-protected with $\varepsilon_1$ and a data set collected at time $t_2$ on a non-disjoint set of individuals is DP-protected with $\varepsilon_2$, the data set obtained by composing the two collected data sets is DP-protected only with $\varepsilon_1+ \varepsilon_2$. Therefore, to attain a certain $\varepsilon$ after $n$ epochs by collecting updates on the same set of clients, each update should be DP-protected with $\varepsilon/n$, thereby exponentially reducing the utility of the data. Given that the guarantees of differential privacy only hold for small values of $\varepsilon$ (below 1), we can conclude that robust DP applied to FL is incompatible with reasonable model accuracy.

Table 2 summarizes, characterizes and compares the privacy attacks and countermeasures discussed so far.

Table 2. Privacy attack methods on FL along with their characteristics and possible countermeasures

| Privacy attack method | Impact | Requirements | Countermeasures | Cost of countermeasures |
|---|---|---|---|---|
| Inference of properties of training data- passive attack (Melis et al., 2019) | The adversary infers properties of the training data without modifying the model | Auxiliary data some of which have the property and some of which do not | • Record-level differential privacy (Shokri and Shmatikov, 2015)<br>• Secure multiparty computation (Bonawitz et al., 2017)<br>• Anonymous communication channels | Low computation but accuracy loss<br>High computation, conflicts with security<br>Medium |
| Inference of properties of training data- active attack (Melis et al., 2019) | The adversary infers properties of the training data by altering the learning procedure | Auxiliary data some of which have the property and some of which do not. The adversary performs local | • Record-level differential privacy (Shokri and Shmatikov, 2015)<br>• Secure multiparty computation (Bonawitz et al., 2017) | Low computation but accuracy loss<br>High computation, conflicts with security |

| | | computations to modify the model | • Anonymous communication channels | Medium |
|---|---|---|---|---|
| Client-side GAN-based attacks (Hitja et al., 2017) | The adversary trains a GAN to generate prototypical samples of the targeted training set | It requires changing the architecture of the distributed model | • Large number of epochs<br><br>• Secure multiparty computation (Bonawitz et al., 2017)<br><br>• Anonymous communication channels | Medium<br><br>High computation, conflicts with security<br><br>Medium |
| FL server-side GAN-based attacks, m-GAN-AI. (Wang et al., 2019) | It enables client-level retrieval of prototypical private data | The adversary operates on the FL server side | • Secure multiparty computation (Bonawitz et al., 2017)<br><br>• Anonymous communication channels | High computation, conflicts with security<br><br>Medium |

## 5. Simultaneously achieving privacy and security in federated learning

Simultaneously offering privacy and security protection in FL while keeping low accuracy loss and low computational overhead is a major challenge. Beyond providing rigorous privacy guarantees, mechanisms employed to protect the privacy of clients should not significantly degrade the accuracy of the learned models. Similarly, privacy-preserving approaches should not drastically increase the computational complexity of the training process or introduce an unaffordable overhead to the network.

Regarding security risks, as discussed in Section 3.3, protection mechanisms against Byzantine and poisoning attacks require the FL server to analyze individual client-supplied updates. Privacy-enhancing techniques based on secure multiparty computation, which use heavy encryption to aggregate local updates before applying them to the global model, do not only entail substantial overhead but they hide the individual updates from the FL server. Therefore, they prevent the server from computing accuracy metrics and weight statistics on individual updates, thus precluding detection and discarding of malicious updates.

Consequently, any technique for protecting clients from privacy attacks should avoid update aggregation in order to be compatible with security protection. As discussed in

Section 4.5, differential privacy fulfills this property and incurs low overhead, but it severely degrades the accuracy of the learned model.

An alternative privacy-enhancing mechanism that, as far as we know, has not been explored yet in the FL literature consists of breaking the link between updates and their generators, e.g., via anonymous communications. To this end, in general it does not suffice to use encrypted communications (SSL/TLS). Even though these ensure the confidentiality of communications against third parties, they are of little help when one of the communicating parties (e.g. the FL server) is a potential adversary. Furthermore, just suppressing identifiers from the message payload may not provide enough privacy guarantees. Metadata of messages, such as the sender's IP address or the timestamp, can reveal private information about individuals and be used by the server to link different updates or pieces of information to single, identifiable individuals. *Anonymous communication channels* are a solution to prevent the communicating parties from learning the metadata of messages, and thus provide sender-message unlinkability.

The rationale of this approach is that, when learning tasks in which the clients' privately held data do not contain personally identifiable information (such as pictures, names, or passport names), breaking the link between clients and their updates is enough to preserve the privacy of clients. The reason is that the privacy sensitivity of non-identifying attributes originates from the risk that profiles are created by an attacker by linking all the values corresponding to the same client; now, such profiling is thwarted if individual updates are made unlinkable.

Unlinkability in communications can be achieved by using:

- Third-party anonymous channels such as proxies, VPNs or TOR (The Onion Routing network[1]).
- Verifiable mixnets.
- Peer-to-peer networks to forward updates among clients before submitting them to the service provider.

The first two approaches require third-party service providers and they are general-purpose off-the-shelf solutions that cannot be tailored to FL. In the case of TOR, it is possible for entry nodes to know who the originators of the messages are, and coalitions of dishonest nodes could reveal the paths of given message transmissions. Additionally, some de-anonymization strategies have been proposed for TOR (Nurmi and Niemelä, 2017). Likewise, a verifiable mixnet requires a third party to host and offer the anonymization service. Thus, both solutions are moving the trust problem to different entities, instead of solving it. Another issue in both approaches is the scalability of both systems in case of broad adoption of privacy-preserving federated learning.

---

[1] https://torproject.org

Let us focus on the third approach, which is the most innovative and can be designed from scratch to encompass the specificities of FL. The basic idea is that, instead of each client always submitting her model updates directly to the FL server (which might leak her private data), she chooses to forward her model update to another client, that we call the forwardee. The forwardee may in turn decide to submit the updated model to the FL server or refuse to do it. If the forwardee refuses the request, the update originator tries another forwardee until someone submits her model update to the FL server. This single-hop approach protects the privacy of the update originator versus the FL server because the latter does not know who originated each received update, and further cannot link successive model updates by the same originator.

The above approach is single-hop and it does not protect the privacy of clients versus each other (the forwardee knows who originated the received update). To fix this, we can generalize the idea into a multi-hop protocol, whereby a forwardee receiving a model update can submit it or forward it to yet another forwardee. Forwarding is an alternative to refusing the updated model. This multi-hop approach prevents privacy disclosure to both FL server and other clients because neither the FL server nor forwardees know whether they are receiving the model update from the originator or from another forwardee. Moreover, the successive model updates originated by the same client are unlinkable, which goes a long way towards guaranteeing that the private data sets of clients stay confidential. Peer-to-peer multi-hop protocols have been successfully applied to enhance privacy in other contexts, such as anonymous query submission to a web search engine (Domingo-Ferrer et al., 2017).

However, for a system like this to be sustainable, clients should be rationally motivated to accept and submit model updates from other clients. A way to introduce artificial incentives in peer-to-peer networks is the use of a decentralized reputation mechanism (Domingo-Ferrer et al., 2016). Briefly, if a client has good reputation (because she has sent updates from others) it will be easier for her to preserve her privacy because she will stand more chances that other clients accept submitting her updates. In this way, clients are motivated to maintain a good reputation by sending model updates from other clients. This creates a virtuous cycle called *co-utility* (Domingo-Ferrer et al., 2017) by which all peers involved are rationally motivated to follow the protocols. Reputation can also be employed to punish clients that have been detected to submit malicious updates. For all this to be possible, privacy-preserving rewarding and punishment protocols should be designed.

The proposed solution is satisfactory in terms of privacy with respect to both FL server and other clients. Unlike differential privacy, model updates are not perturbed and thus the accuracy of the model is entirely retained. Moreover, the fact that model updates are neither perturbed nor aggregated allows the FL server to compute statistics on the weights and biases of the client updates in order to detect and discard defective (or malicious) updates. The downside is that peer-to-peer forwarding of updates requires clients to communicate with each other and, therefore, introduces a communication overhead. However, the computational cost incurred by this approach is significantly lower than that of secure aggregation protocols based on SMC, which employ heavy cryptography.

## 6. Conclusions

Federated learning is vulnerable to Byzantine and model poisoning attacks by design. The implications are either i) failure to converge to an optimal model or ii) vulnerability to the introduction of adversarial learning targets. Fortunately, our experiments have shown that countermeasures against security attacks are effective at thwarting attacks conducted by large coalitions of malicious clients. Among the discussed and tested countermeasures, the Krum aggregation rule offered the best results, both in terms of effectiveness at discarding malicious updates and of low complexity.

However, all these countermeasures require the FL server to access individual updates from the clients. Thus, the protection of the clients' privacy via the use of cryptographic aggregation mechanisms makes the detection of potentially malicious clients impossible. Even though $\epsilon$-differential privacy allows protecting the privacy of clients while still allowing server access to individual updates, the distortion added to the updates severely degrades the accuracy of the global model to unacceptable levels.

To tackle the limitations of these privacy-preserving methods (and, therefore, to provide security, privacy and accuracy in FL) we have proposed to break the link between clients and their updates using anonymous communication channels enforced via peer-to-peer decentralized networks, as we discussed in Section 5. These allow detection of potentially malicious clients while protecting client privacy and preserving the accuracy of the model. The challenges to be tackled in order to bring this solution into practice include i) the design of update forwarding and submission protocols, ii) the incorporation of artificial incentives, i.e., reputations, to motivate peers to collaborate according to the protocols and, iii) the design of privacy-preserving protocols to reward good updates and punish bad updates.

## Acknowledgements

We gratefully acknowledge support from Huawei Technologies Oy (Finland) (agreement no YBN2019035188), the European Commission (projects H2020-871042 "SoBigData++" and H2020-101006879 "MobiDataLab"), the Government of Catalonia (ICREA Acadèmia Prize to J. Domingo-Ferrer and grant 2017 SGR 705), and the Spanish Government (projects RTI2018-095094-B-C21 "Consent" and TIN2016-80250-R "Sec-MCloud"). The authors from URV are with the UNESCO Chair in Data Privacy, but the views in this paper are their own and are not necessarily shared by UNESCO.